# Transition from confined to bulk dynamics in symmetric star–linear polymer mixtures

*Daniele Parisi,[1,2] Domenico Truzzolillo,[1,3] Vishnu D. Deepak,[4] Mario Gauthier,[4] Dimitris Vlassopoulos[1,2]*

[1]FORTH, Institute of Electronic Structure and Laser, University of Crete, 70013 Heraklion, Crete, Greece

[2]Department of Materials Science and Technology, University of Crete, 70013 Heraklion, Crete, Greece

[3]Laboratoire Charles Coulomb (L2C), UMR 5221 CNRS Université de Montpellier, Montpellier, France.

[4]Department of Chemistry, University of Waterloo, Waterloo, Ontario N2L 3G1, Canada

**ABSTRACT**

We report on the linear viscoelastic properties of mixtures comprising multiarm star (as model soft colloids) and long linear chain homopolymers in a good solvent. In contrast to earlier works, we investigated symmetric mixtures (with a size ratio of 1) and showed that the polymeric and colloidal responses can be decoupled. The adopted experimental protocol involved probing the linear chain dynamics in different star environments. To this end, we studied mixtures with different star mass fraction, which was kept constant while linear chains were added and their entanglement plateau modulus ($G_p$) and terminal relaxation time ($\tau_d$) were measured as functions of their concentration. Two distinct scaling regimes were observed for both $G_p$ and $\tau_d$: at low linear polymer concentrations, a weak concentration dependence was observed, that became even weaker as the fraction of stars in the mixtures increased into the star glassy regime. On the other hand, at higher linear polymer concentrations, the classical entangled polymer scaling was recovered. Simple scaling arguments show that the threshold crossover concentration between the two regimes corresponds to the maximum osmotic star compression and signals the transition from confined to bulk dynamics. These results provide the needed ingredients to complete the state diagram of soft colloid–polymer mixtures and investigate their dynamics at large polymer-colloid size ratios. They also offer an alternative way to explore aspects of the colloidal glass transition and the polymer dynamics in confinement. Finally, they provide a new avenue to tailor the rheology of soft composites.

## I. INTRODUCTION

It is known that mixing polymers with nanosized particles can lead to novel materials with enhanced physico-chemical properties, for example superior strength, improved processability, reduced permeability and decreased rolling resistance in tires.[1–3] Recently, the established picture[4–6] of entanglement dynamics in long linear polymers has been invoked to rationalize the dynamics of nanocomposites. Simulations have shown that, as a consequence of chain confinement due to the presence of non-interacting nanoparticles, the entanglement density can be reduced.[7] It was indeed shown that nanoporous materials under strong confinement led to enhanced entanglement of a linear chain matrix. [8–10] While understanding polymer chain dynamics in the presence of confinement remains a challenge with both scientific and technological implications,[11–14] colloid–polymer mixtures have inexorably emerged as a paradigm to address the entropic manipulation of the flow properties of soft matter and the colloidal glass transition.[15–17] Soft colloids such as star polymers, microgels, core–shell particles and star-like micelles offer further opportunity to tune the “softness”[15,16,18,19] of mixtures, by changing their molecular architecture and for this reason, their usage opens new perspectives for the formulation of novel soft materials in fields ranging from chemicals to foodstuff and medical applications. The addition of linear chains to soft colloidal suspensions is considered to be a facile and effective way to tailor the stability of nanoparticles,[20] the flow and the microscopic dynamics at rest of colloids.[11,15,21–27] In contrast to hard colloids, when chains are added to suspensions of soft particles, the latter may deform and compress due to the osmotic pressure exerted by the chains and by the colloids themselves and their low elastic modulus.[15] Hence the problem of linear chain confinement becomes more complex, being dependent on the elastic modulus of the particles. In particular, while for binary long chain polymer–hard colloid nanocomposites the free chain dynamics are uniquely driven by the competition between the mesh size (tube diameter) and the average distance between the hard colloids, in ternary solvent–soft colloid–polymer mixtures, colloid deformability and penetrability play a crucial role in determining polymer relaxation: the transition from bulk to confined dynamics of linear chains becomes sensitive to the subtle interplay between the excluded volume and the configurational entropic part of the colloid free energy,[28] the latter being dependent on the internal degrees of freedom and topology of the colloid, and the osmotic pressure exerted by the chains on the colloids.

We have already shown in a previous work[29] that linear and star polymer rheology can be decoupled, provided that the two species (linear chains and stars) exhibit two well-separated relaxation times. Star–linear polymer mixtures are useful model systems to investigate free polymer chain dynamics in mixtures in the absence of enthalpic interactions, since the soft colloidal stars are osmotically affected by the addition of linear polymers. Stars are representative of a large class of long hairy particles, including block copolymer micelles and grafted colloids, where the polymeric nature of the hairs has a stabilizing effect and determines to a large extent their macroscopic response. At large volume fractions, they exhibit a glasslike transition[30–32] whose main features are an enhanced frequency-independent storage modulus (G'), which is much larger than the loss modulus (G"), and a non-ergodic plateau in the intermediate scattering function.[31,32] Many intriguing and non-trivial phenomena characterize the dynamics of star–linear polymer mixtures. It was observed that small amounts of added linear homopolymer with a size smaller than the star leads to glass melting due to depletion.[21,31–34] This phenomenon is akin to that widely studied in hard colloid–polymer mixtures,[16,35] though bearing a distinct feature: the osmotic force due to the added polymers can squeeze the stars, yielding a size reduction and, at high concentrations, star aggregation and possibly the microphase separation of collapsed stars.[29,31,34,36] It has important consequences not only on colloid dynamics but also on the self-assembly of micelles.[37] Note however that nearly all previous work considered mixtures with linear-to-star size ratios well below 1, hence linear polymers with relatively low molecular weights, which severely limited the detection of the viscoelastic response of the polymer matrix. For this reason, high molecular weight polymers, that form entanglements more easily, are ideal candidates to probe linear chain dynamics when they are mixed with colloids of the same size, and to shed light on the role of soft confinement on their terminal relaxation. Moreover, by employing chains whose size is comparable to that of the stars is important for completing the general description of soft colloid-polymer mixtures.[15] Importantly, in such a case the size of polymers is larger than the average distance between the outer blobs of the stars and polymer–colloid interpenetration is reduced; in other words, wetting of the colloids by the linear chains is limited and so is its influence on osmotic shrinkage of the stars.

All in all, whereas the main features of colloidal dynamics in this kind of asymmetric soft composite systems have been investigated in detail, the dynamics of the linear chains has not been explored, which leaves outstanding challenges concerning the role played by soft confinement on the rheology of polymers and in general the dynamic of symmetric soft colloid-polymer mixtures. In particular, the questions we wish to address in this work are: *(i) under which conditions (polymer and colloid volume fractions) is a transition from confined to bulk dynamics of linear chains observed in soft colloid–polymer mixtures, and which are its rheological signatures? (ii) can colloid and polymer dynamics in these mixtures be decoupled, and can the onset of this transition be determined?* The aim of this work is to elucidate the influence of soft confinement produced by the inclusion of star polymers on the dynamics of entangled polymers. To this end, the viscoelastic response of symmetric star–linear polymer mixtures was investigated in a nearly good solvent, where star and linear dynamics can be decoupled. By varying the mixture composition, two distinct relaxation regimes can be identified, which allows the determination of the effect of stars on the entanglement dynamics of free chains. The first regime, which is more pronounced at low linear polymer concentrations, is governed by star-induced confinement, where the disentanglement time and plateau moduli of the polymer matrices exhibit a rather unprecedented behavior: a weak concentration dependence is observed at low linear polymer concentrations, which further weakens as the star concentration is increased. As the linear polymer concentration is increased further, a second scaling regime eventually emerges where the classical scaling of entangled polymer solutions is observed. The crossover between these two regimes defines the confined-to-bulk transition of chain dynamics in the mixtures and is affected by (and affects) the stars, that shrink under the influence of the osmotic pressure exerted by the surrounding chains. These novel findings are discussed in detail below.

## II. EXPERIMENTAL

### II.1. Materials

The multiarm polybutadiene (PBD) star used in the investigation, identified as S362, contained more than 90% 1,4-butadiene units, had a number-average branching functionality $f_n$ = 362 arms (weight-average branching functionality $f_w$ = 392), a weight-average molar mass $M_w^s = 9.8\times10^6$ g/mol, and a polydispersity $M_w/M_n$ = 1.14. Each arm had $M_w^a$ = 24400 g/mol and $(M_w/M_n)^a$ = 1.06.[38] Details on the synthesis and the size exclusion chromatography (SEC) analysis of that material were reported elsewhere.[38,39] According to the Daoud-Cotton model,[40] a star polymer in a good solvent is characterized by a non-homogeneous monomer density distribution that comprises three regions: the inner melt-like core, the intermediate ideal region and the outer excluded volume region. The latter is involved in interactions with neighboring stars in crowded suspensions. The linear PBD used, identified as L1000, was obtained from Polymer Source (Canada) and had a weight-average molar mass $M_w^L$ = 1060000 g/mol and a polydispersity $M_w/M_n$ =1.1. The polymers were dissolved in squalene, a relatively good (see below) and non-volatile solvent.[41] The hydrodynamic radii, determined through dynamic light scattering (DLS) measurements in dilute solution at 20 °C, were $R_h^s$ = 39 nm and $R_h^L$ = 41 nm (see Supplemental Information (SI), Figures S1 and S2), yielding a L/S size ratio close to 1. The respective overlap concentrations were $C_s^*$ = 60.6 mg/ml and $C_L^*$ = 6.19 mg/ml. Five pure star polymer suspensions were prepared at different effective volume fractions $\varphi_s = C_s/C_s^*$ within a range $0.5\leq\varphi_s\leq4.0$. In the absence of linear chains, star suspensions vitrify at $\varphi_s = \varphi_{Gs}$, with $1.5<\varphi_{Gs}<2.0$, i.e., the volume fraction where the structural relaxation time becomes larger than 100 s, time at which several authors assign the nonergodicity transition.[42–44] It should be noted that when these suspensions are out of equilibrium, they exhibit time-dependent dynamics (aging)[45–47] which can be taken into account (see Figure S3 in the SI). For $\varphi>\varphi_{Gs}$ star polymer suspensions are viscoelastic solids, with both storage (G') and loss (G") moduli weakly frequency-dependent, G'>G", and G" exhibiting a shallow, broad minimum typical of glassy colloids.[32,47–50] When preparing mixtures with linear polymers, the same mass fraction of star polymers was maintained, i.e., the added linear chains replaced part of the solvent. However, complete solvent removal was never reached.

**II.2. Rheology**

The dynamics of the star–linear mixtures were investigated through rheological measurements with a sensitive strain-controlled rheometer (ARES-HR 100FRTN1 from TA, USA). Due to the very limited amounts of samples available, a small home-made cone-and-plate geometry (stainless steel cone with 8 mm diameter and a 0.166 radians cone angle) was mostly used. At very low concentrations, a 25 mm stainless steel cone (with angle equal to 0.02 radians) was used to increase the torque signal. The temperature was set to 20.00 ± 0.01 °C and controlled using a Peltier plate with a recirculating water/ethylene glycol bath. During an experimental run the sample (which had a pasty appearance) was loaded on the rheometer, and a well-defined pre-shear protocol was applied such that each sample was subjected to: (i) a dynamic strain amplitude sweep at fixed frequency (100 rad/s) to determine the linear viscoelastic regime, i.e., where the moduli did not show any detectable dependence on strain amplitude; (ii) a dynamic time sweep at large nonlinear strain amplitude (typically 200%) and low frequency (1 rad/s), to shear-melt (fully rejuvenate) the sample, as judged by the time-independent first harmonics $G'(\omega,\gamma_0)$ and $G''(\omega,\gamma_0)$ (this step typically lasted 300 s); (iii) a dynamic time sweep for a (waiting) time $t_w \approx 10^5$ s, which was performed in the linear regime to monitor the time evolution of the moduli to steady state, corresponding to an aged sample; (iv) small-amplitude oscillatory shear (SAOS) tests within the frequency range 0.01-100 rad/s, to probe the linear viscoelastic spectrum of the aged samples. The data shown hereafter refer to aged samples, and the influence of aging will not be discussed further.

## III. RESULTS AND DISCUSSION

### III.1. LINEAR VISCOELASTICITY (LVE)

The LVE spectra for S362 suspensions in the absence of linear chains are provided in Figure 1-A. At $\varphi_s = 2.0$ and $\varphi_s = 4.0$, the stars exhibit typical colloidal glassy dynamics over four decades of frequency. Such a solid-like behavior is characteristic for aged suspensions.[39,45,47] Note that no systematic rheological investigation or light scattering characterization of the pure star suspensions (as done, for example, by Pellet and Cloitre for microgels[51]) was carried out to identify possible distinct glassy and jammed regimes in the star polymers, as this goes beyond the scope of this work.

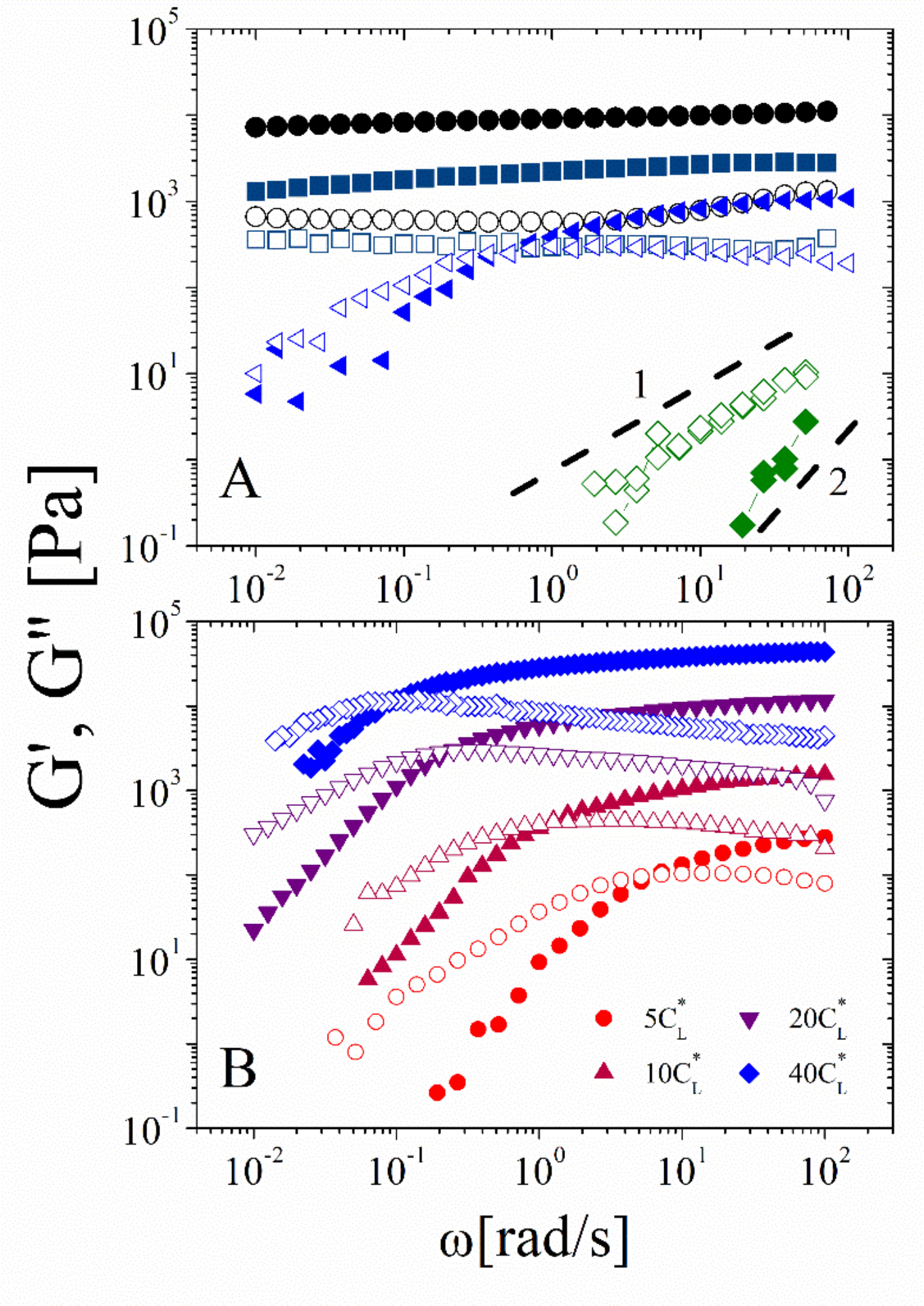

Figure 1. Panel A: G'(ω) (filled symbols) and G"( ω) (empty symbols) in the linear viscoelastic regime ($\gamma_0$<0.5%) for pure star suspensions at $\varphi_s$ = 0.9 (green lozenges), 1.5 (triangles), 2.0 (squares), $\varphi_s$ = 4.0 (circles). Panel B: G'(ω) (filled symbols) and G"( ω) (empty symbols) in the linear viscoelastic regime ($\gamma_0$<0.5%) for pure linear polymer solutions at different concentration as shown in the panel.

For lower fractions ($\varphi_s$ = 0.9) the S362 suspensions exhibit a response typical for a viscoelastic liquid, with G"(ω)>>G'(ω) and respective frequency scaling of 1 and 2, whereas G'(ω) is not resolved at the lowest frequencies where the stress $\sigma$ = G"(ω) $\gamma_0$ is entirely dictated by the out-of-phase response of the system (Figure 1-A). For even lower star polymer fractions ($\varphi_s$<0.9), the response is entirely dominated by its viscous component (Newtonian response) and no storage modulus can be measured.

Some of these pure star polymer samples were selected to investigate the rheology of symmetric star–linear polymer mixtures as discussed below. Figure 1-B depicts selected LVE spectra for pure L1000 solutions at different effective volume fractions $5 \leq \varphi_L \leq 40$, where $\varphi_L$, similarly to the star polymers, is defined as $C_L/C_L^*$. It is worth pointing out that the linear polymer concentration, in both mixtures and pure solutions (without stars), is expressed as the nominal concentration (mg/ml) of chains excluding the stars, i.e. $C_L = \frac{W_L}{\left(V_{sol}+\frac{W_L}{\rho_L}\right)}$, where $W_L$ and $\rho_L$ are the mass and the density of the dissolved chains and $V_{sol}$ is the volume of small molecule solvent (squalene) in each sample.

As expected, the linear polymer chains exhibit a continuous slowdown of the dynamics, as evidenced by the progressive shift of the crossover between the two moduli to lower frequencies and an increase in the plateau modulus for increasing polymer concentrations. In the entanglement regime, such a behavior has been largely discussed and accurately described in the literature based on reptation,[6,52–54] and scaling predictions for both the terminal time and the plateau modulus under different solvency conditions are available. In fact, it should be remembered that the volume fraction dependence of the plateau modulus follows a power law with an exponent of 2.3 under both athermal and theta conditions, whereas the disentanglement time exhibits power law dependence with exponent values of 2.14 and 2.87 under good and theta conditions, respectively.[53,54] Moreover, for L1000 solutions no aging has been observed, as expected. Hence, the two pure components of the S362/L1000 mixtures exhibit dramatically different concentration dependencies of their linear viscoelastic response, reflecting different relaxation mechanisms. Such a dichotomy of the dynamics facilitates probing of the chain dynamics in the mixtures.

The dependence on L1000 concentration of selected LVE spectra for star/linear mixtures is illustrated in Figure 2 at four different S362 effective volume fractions. All the spectra are characterized by one common feature that is absent in the pure S362 suspensions: G"(ω) exhibits a local maximum whose position shifts to lower frequencies as the L1000 content increases (shaded connected circles in Figure 2). At the same time, G′(ω) decreases monotonically with decreasing frequency and becomes increasingly frequency-dependent at higher L1000 contents. The data for S362 suspensions at $\varphi_s$ = 2.0 and $\varphi_s$ =4.0, depicted in Figures 2-C-D, indicate the appearance of a low-frequency plateau, clearly distinct from the higher-frequency plateau associated with L1000. The former is due to the slowest component in the mixtures, i.e. the glassy stars. Hence, the linear dynamics of the star and linear polymers are unambiguously decoupled and can be investigated in detail. It is also worth mentioning that the mixture at $\varphi_s$ = 0.9 (Figure 2-B) displays the same low-frequency plateau, indicating the unexpected vitrification upon adding linear chains of an initially fluid star suspension. The description of such a phenomenon is out of the scope of this work and will not be detailed further hereafter.

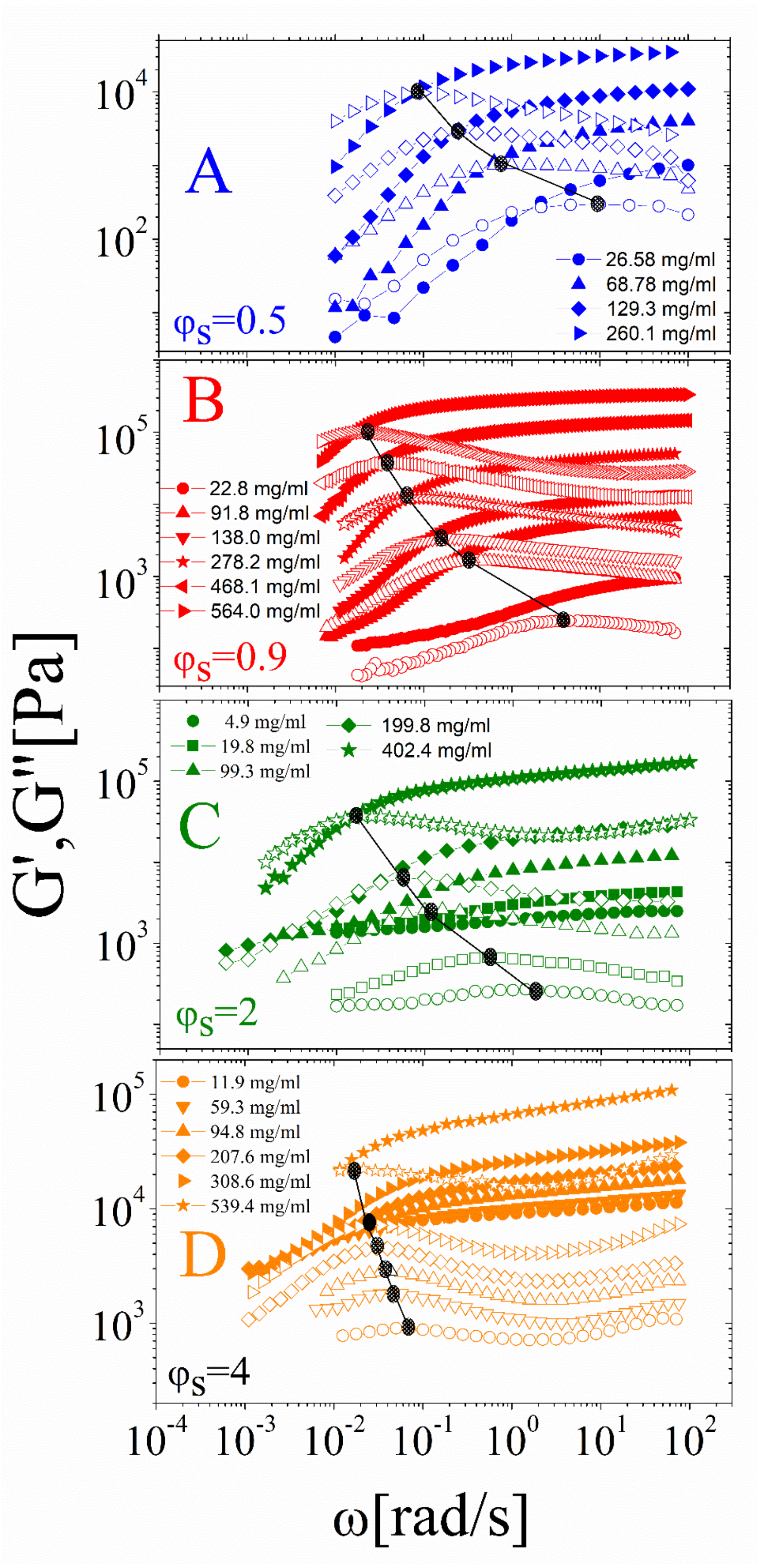

Figure 2. Selected LVE spectra for different L1000–S362 mixtures. A) $\varphi_s = 0.5$, B) $\varphi_s = 0.9$, C) $\varphi_s = 2.0$, D) $\varphi_s = 4.0$. The shaded connected circles indicate the maximum G" observed in each LVE spectrum and the evolution of the L1000 relaxation time.

To explore the polymer disentanglement dynamics, the longest relaxation time, $\tau_d$, was determined as the inverse of the frequency at the local maximum G"($\omega_m$), $\tau_d = 1/\omega_m$ (Figure 2),[53] and its dependence on the L1000 concentration was monitored. The exact position of the maximum G"($\omega_m$) was extracted from the LVE spectrum by fitting the loss modulus around the maximum with a parabola, avoiding the use of a specific model.

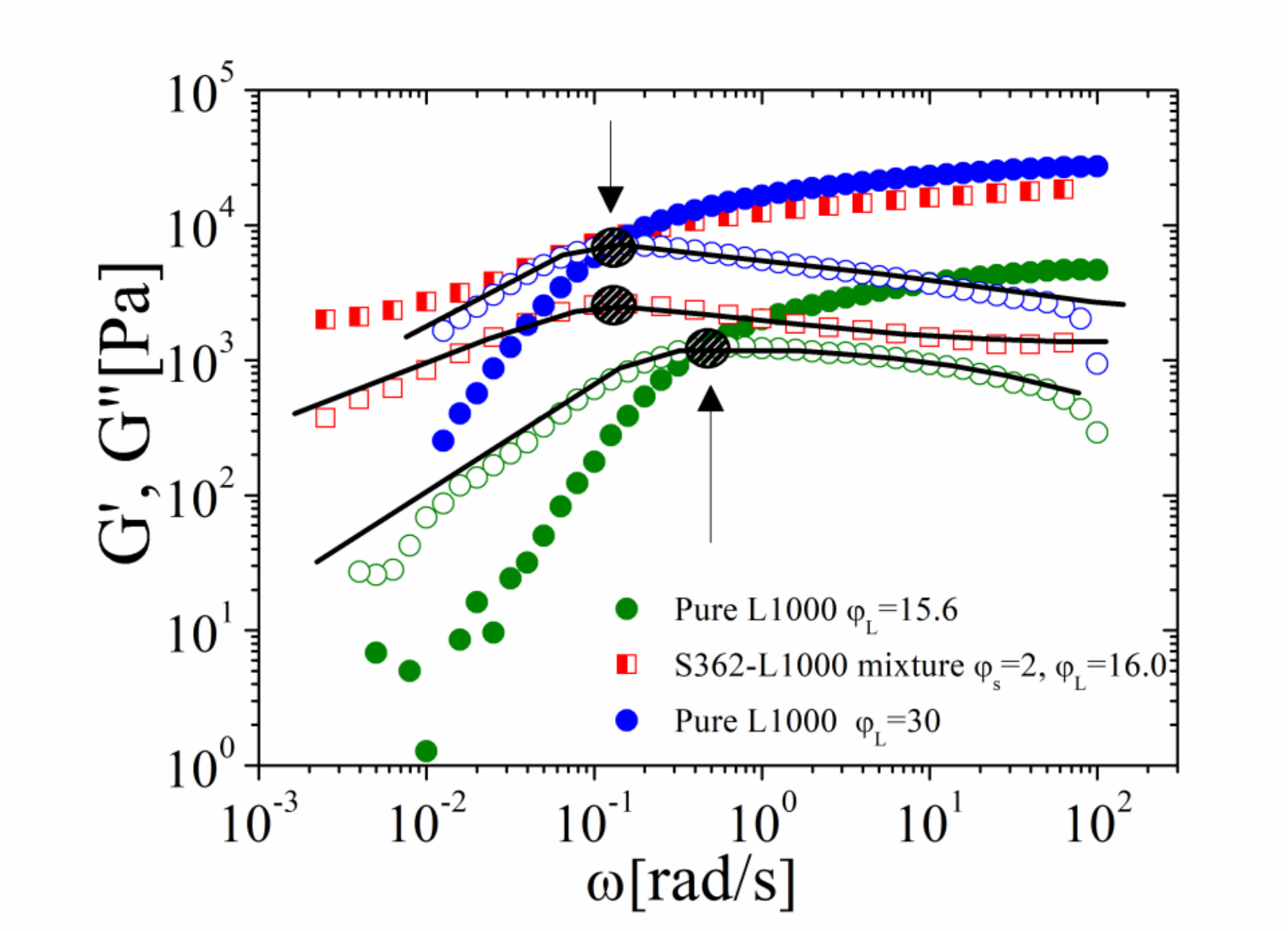

Figure 3. Determination of the L1000 relaxation time by matching the terminal times for pure L1000 solutions and a S362–L1000 mixture. The longest relaxation time in a S362–L1000 mixture is longer than in a pure L1000 solution at the same L1000 content (here $\varphi_L \approx 16$). The same relaxation time of the S362–L1000 mixture is recovered at a higher L1000 concentration, here $\varphi_L$ =30. The arrows and the dashed green circles show the position of the maximum in G". The solid lines are drawn to guide the eye.

At a sufficiently low L1000 concentration, a shift in the relaxation time for the entangled linear polymer $\tau_d$, caused by the inclusion of star polymer, is clearly distinguishable (Figure 3). At a set value of $C_L$ concentration, calculated by excluding the volume occupied by the stars, the local maximum in G" is systematically shifted to lower frequencies when star polymers are present: linear chains in the mixtures have access to a lower volume with respect to the star-free solutions, thus behaving like more concentrated solutions. It has been shown[29] that such a shift cannot be properly quantified without accounting for osmotic shrinkage of the stars induced by the linear matrix (osmotic de-swelling). Using this concept, it will be shown below through a simple scaling argument that the observed shift in the terminal relaxation time is indeed compatible with star de-swelling that depends on both the L1000 and S362 fractions in the mixtures. Such a shift becomes gradually less pronounced as $C_L$ increases, i.e. when the tube diameter characterizing the L1000 matrix becomes comparable to or even smaller than the apparent surface-to-surface distance between the cavities of size R containing the stars. In other words, the dominant confinement length becomes dictated by the linear chains, while confinement due to the stars becomes less effective or even negligible. Concomitantly, it can be seen that the high frequency storage modulus of an S362–L1000 mixture matches that of a pure L1000 solution with the same $\tau_d$, corroborating the scenario that the high frequency plateau modulus is dominated by the confined L1000 matrix (Figure 3, where plateau moduli of pure L1000 and mixture are within a factor of 2).

The extracted $\tau_d$ values for the pure L1000 solutions and the S362–L1000 mixtures are summarized in Figure 4, along with those obtained for pure L1000 semidilute suspensions ($\varphi_s$ = 0), as a function of the L1000 concentration. The scaling relation $\tau_d \sim C_L^{2.5\pm0.2}$ is observed[53,54] for pure L1000. Using the predicted dependence of the longest relaxation time for semidilute entangled linear polymers $\tau_d \sim C_L^{(3.4-3\nu)/(3\nu-1)}$, the Flory exponent for the PBD chains in squalene, $\nu$ = 0.56±0.01, can be extracted. This confirms that squalene is a nearly good solvent for PBD.

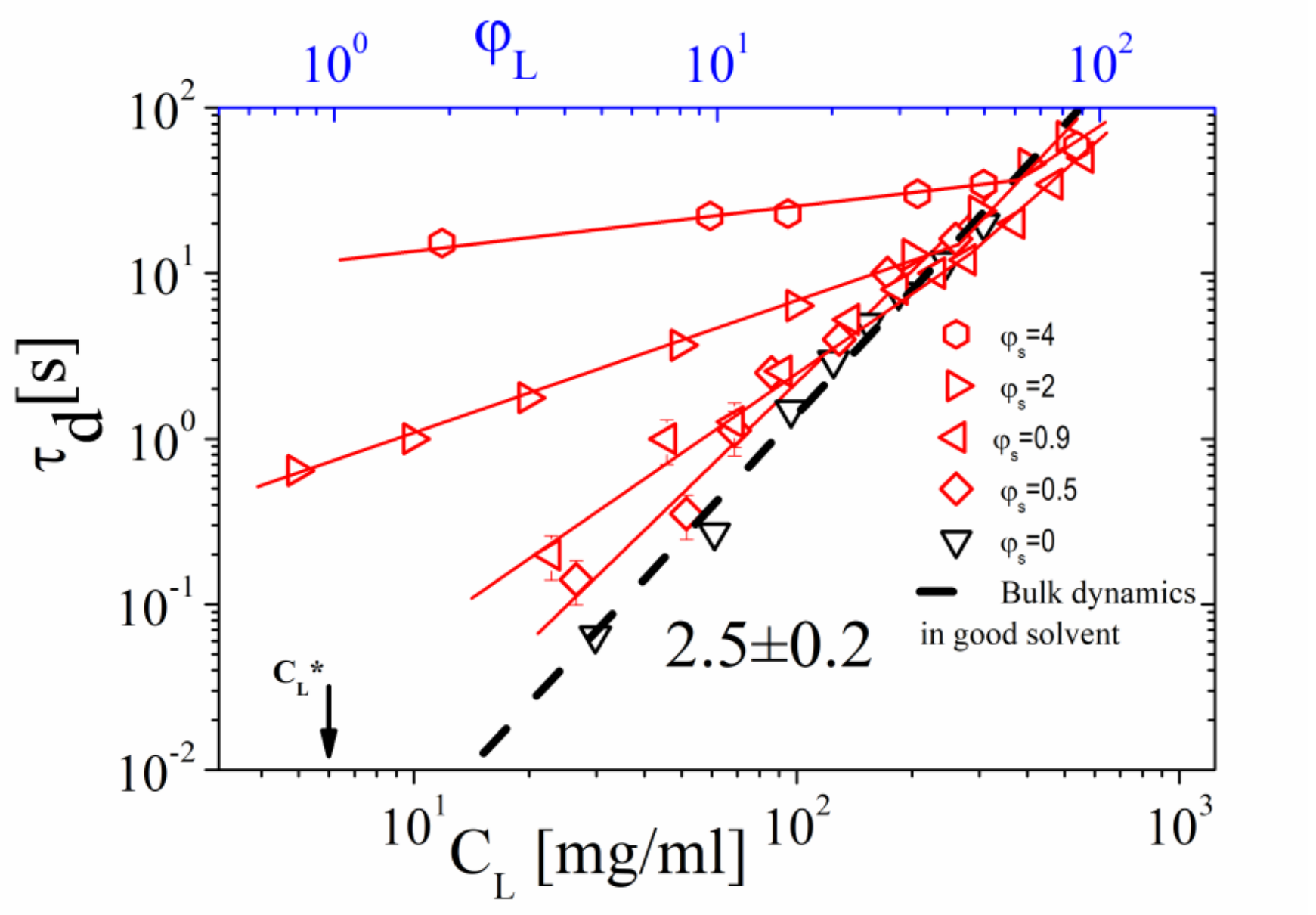

Figure 4. Disentanglement time $\tau_d$ as a function of the concentration of linear chains and the volume fraction of star polymer. The black dashed line represents the obtained power-law concentration dependence of $\tau_d$ for pure linear chains in squalene. The red lines are just a guide for the eye. The top x-axis shows the effective volume fraction of the linear chains. The black arrow indicates the overlap concentration of the linear chains (6.19 mg/ml).

At the lowest S362 effective volume fraction ($\varphi_s$ = 0.5), the effect of confinement on linear chain dynamics is barely detectable: $\tau_d$ in the S362–L1000 mixture only has a slightly higher value and exhibits a slightly weaker dependence on $C_L$ as compared to $\tau_{d,pure}$. However, upon increasing the star volume fraction ($\varphi_s \geq 0.9$), a clear deviation from the $\tau_{d,pure}$ scaling is observed in the low $C_L$ regime, where the stars are partially swollen, since osmotic de-swelling due to the L1000 matrix is negligible (see below). The effective blob density of the L1000 matrix in the mixtures is larger than their nominal one, i.e. the blob density of a solution with the same linear polymer-to-solvent mass ratio in the absence of stars. Consequently, the terminal (disentanglement) time $\tau_d$ is larger than in a pure L1000 solution at the same $C_L$: linear chains in the S362–L1000 mixtures are confined. On the other hand, as $C_L$ in the mixtures is increased, the terminal time $\tau_d$ progressively increases until eventually crossing over to the behavior of pure L1000 solutions, as indicated by the data for $\varphi_s = 2$ and (marginally) $\varphi_s = 4$. It is worth pointing out that this phenomenology (sketched in a simplified form in Figure 5), including the exact recovery of pure L1000 entanglement dynamics at high $C_L$, holds true especially when there is no substantial entanglement dilution that would speed-up L1000 relaxation.[55] This is because of the osmotic compression (de-swelling) of the stars which is discussed below. Note also that there is no reinforcement due to filler inclusion as reported for nanocomposites.[56,57] The mechanisms at work when hard fillers are added to polymer matrices are not accounted for by dynamics dictated uniquely by the effective volume accessible to the chains, i.e. by their configurational space. Therefore, at large $\varphi_s$ values, the disentanglement time $\tau_d$ and the plateau modulus $G_p$, while converging toward the same power-law scaling observed for pure L1000 solutions, may not fully coincide with $\tau_{d,pure}$ and $G_{p,pure}$. As will be shown below, an analysis based on the assumption of dynamics dictated simply by the configurational space of the chains and on star shrinkage conforms well the experimental data.

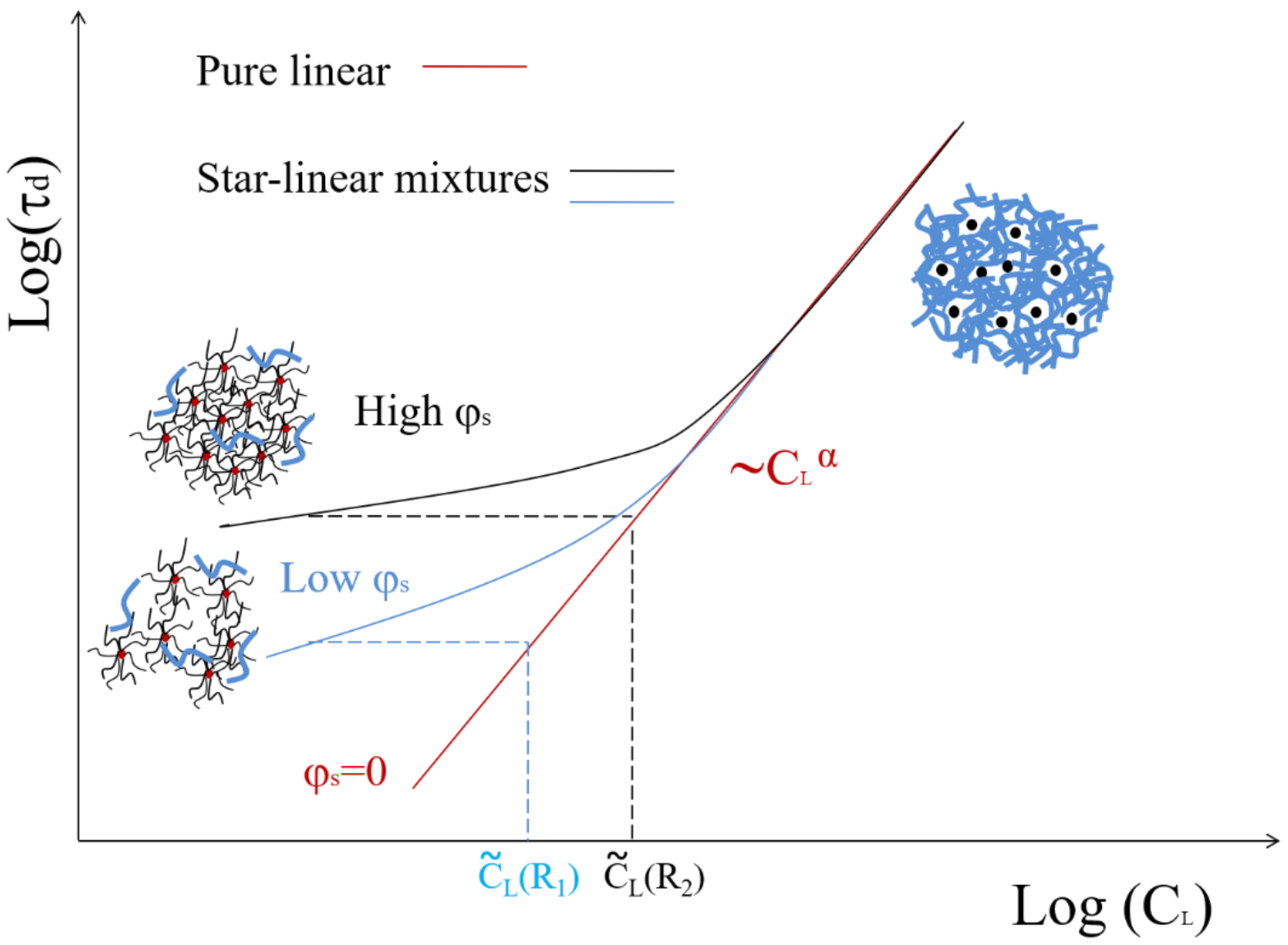

Figure 5. Schematic illustration of dynamic mapping allowing the estimation of the effective linear polymer concentration in the S362–L1000 mixtures. Each $\tau_d$ measured for a given S362–L1000 mixture is mapped (dashed horizontal lines) into a corresponding value of the power law scaling line for the pure L1000 solution. The extracted respective concentration $\tilde{C}_L(R)$ is the effective L1000 concentration in the S362–L1000 mixtures used to determine the size R of the stars.

Indeed, if the effective volume fraction of the star S362 is low enough, in other words if i) the linear polymer in the mixture enters the semidilute regime for a finite linear chain content well before reaching the melt state, ii) the linear polymer is depleted from the interior of the stars and fully wetted by the molecular solvent, and iii) the stars shrink and leave a positive interstitial free volume, such a behavior can be rationalized if we consider linear chains confined by spherical soft cavities (stars) of radius R, partially wetted by the solvent. The concentration $\tilde{C}_L(R)$ of the L1000 chains under such star confinement can be calculated as:

$$\tilde{C}_L(R) \approx \frac{W_L}{V_{sol} - N_s \frac{4}{3}\pi R^3 + \frac{N_s M_w^s}{N_A \rho_s} + \frac{W_L}{\rho_L}} \quad (1)$$

where $W_L$ is the mass of chains dissolved in the mixture, $V_{sol}$ is the solvent (squalene) volume, $N_s$ is the number of stars, $\rho_s$ and $\rho_L$ are the density of the S362 and L1000 components, respectively; the values $\rho_s = \rho_L$ = 892 mg/ml, corresponding to the density of poly(1,4 butadiene), were used in the current case. It is worth recalling the meaning of each term in the denominator of Equation (1): the second term represents the volume inaccessible to linear chains, i.e., the volume of all the spherical obstacles (stars) in the mixture; the third and the fourth terms are the bulk volume of the stars and the linear chains, i.e., the volume occupied by the polymer phase in the absence of solvent. It should be remembered that $\tilde{C}_L(R)$ differs from the control variable $C_L$ used in the experiments: the former represents the concentration of chains excluding the volume of the spherical cavities (stars), but with a portion of the solvent, as expressed by Equation (1), whereas the latter is the nominal concentration of chains excluding the stars, $C_L = W_L/(V_{sol} + W_L/\rho_L)$.

Therefore, Equation (1) can be used to understand to a first approximation the crossover of mixture dynamics towards that of L1000 solutions. As a matter of fact, if Equation (1) is modified by introducing the bulk radius of the stars in their completely collapsed state,

$$R = R_c = \left( \frac{3M_w^s}{4\pi N_A \rho_s} \right)^{1/3} \quad (2)$$

the following equation is obtained

$$\tilde{C}_L(R) = C_L = \left( \frac{W_L}{V_{sol} + \frac{W_L}{\rho_L}} \right)^{1/3} \quad (3),$$

which represents the L1000 concentration in star-free suspensions. Hence, from Equations (1–3), it can be inferred that at high $C_L$, when the stars effectively release the molecular solvent (squalene) from their interior due to the osmotic pressure exerted by the linear chains, the dynamics must converge to that of pure L1000 solutions. Although this ensures that the scaling observed in linear polymer solutions is recovered at sufficiently high L1000 contents, it does not guarantee, as previously mentioned, that the disentanglement times $\tau_d$ and $\tau_{d,\ pure}$ coincide. This is because, within the above framework, all the effects associated with a possible speeding-up or slowing-down of the polymer matrix relaxation (and respective weakening or strengthening of its modulus), due to the inclusion of stars, are neglected. Indeed, for most of the mixtures investigated (see Figures 4 and 7) the same power-law scaling is recovered. Nevertheless, some of the data suggest a possible weak speeding-up of the dynamics and weakening of matrix modulus (lower values of $\tau_d$ and $G_p$, respectively) at $C_L$ above the star polymer collapse threshold, where the relaxation times and moduli scale similarly to pure semidilute polymer solutions (see $\tau_d$ at $\varphi_s = 0.9$ and $G_p$ for $\varphi_s = 4.0$). This marginal effect is outside the thrust of this work and, given the limited amount of data available, it will not be discussed further.

Equation (1) was used to determine the size of the stars in the mixtures according to the protocol represented in Figure 5: i) the data for the pure linear polymers were fitted with a power-law function $\tau_d = KC_L^{\alpha}$, yielding $K = (1.4\pm0.5)10^{-5}$ and $\alpha = 2.5\pm0.1$. This allows estimating $\tau_d$ for the star-free solutions when $C_L$ lies within the semidilute regime; ii) For each mixture, the respective $\tau_d$ was mapped onto the power-law $\tau_d = KC_L^{\alpha}$ as described in Figure 5; and iii) The extrapolated value of $\tilde{C}_L(R)$, i.e., the effective concentration of linear chains in pure L1000 solutions having the same $\tau_d$ as the mixtures was calculated (see Figure 5). Such mapping is meaningful only if the chain dynamics are dominated by chain–chain rather than chain–star entanglements, the latter being relevant when the stars are highly swollen. This hypothesis is easily tested by computing the size of the spherical and "impenetrable" part of the stars R. Indeed, by solving Equation (1) using the extrapolated value of $\tilde{C}_L(R)$, the average size of the spherical cavities (stars) confining the linear chains can be determined. The values of $R(C_L)$ obtained for $\varphi_s = 0.9$, $\varphi_s = 2$ and $\varphi_s = 4$ (Figure 6) demonstrate that the contribution of the stars to the overall response of the mixtures is significant, as judged by the decrease in R with $C_L$. This shrinkage of the stars upon adding linear chains is expected from simple osmotic considerations.[29,39,58] For all the mixtures, $R$ decreases from a value $R_0$ (L1000-free suspensions) to a value $R_c$ corresponding to the collapsed stars.

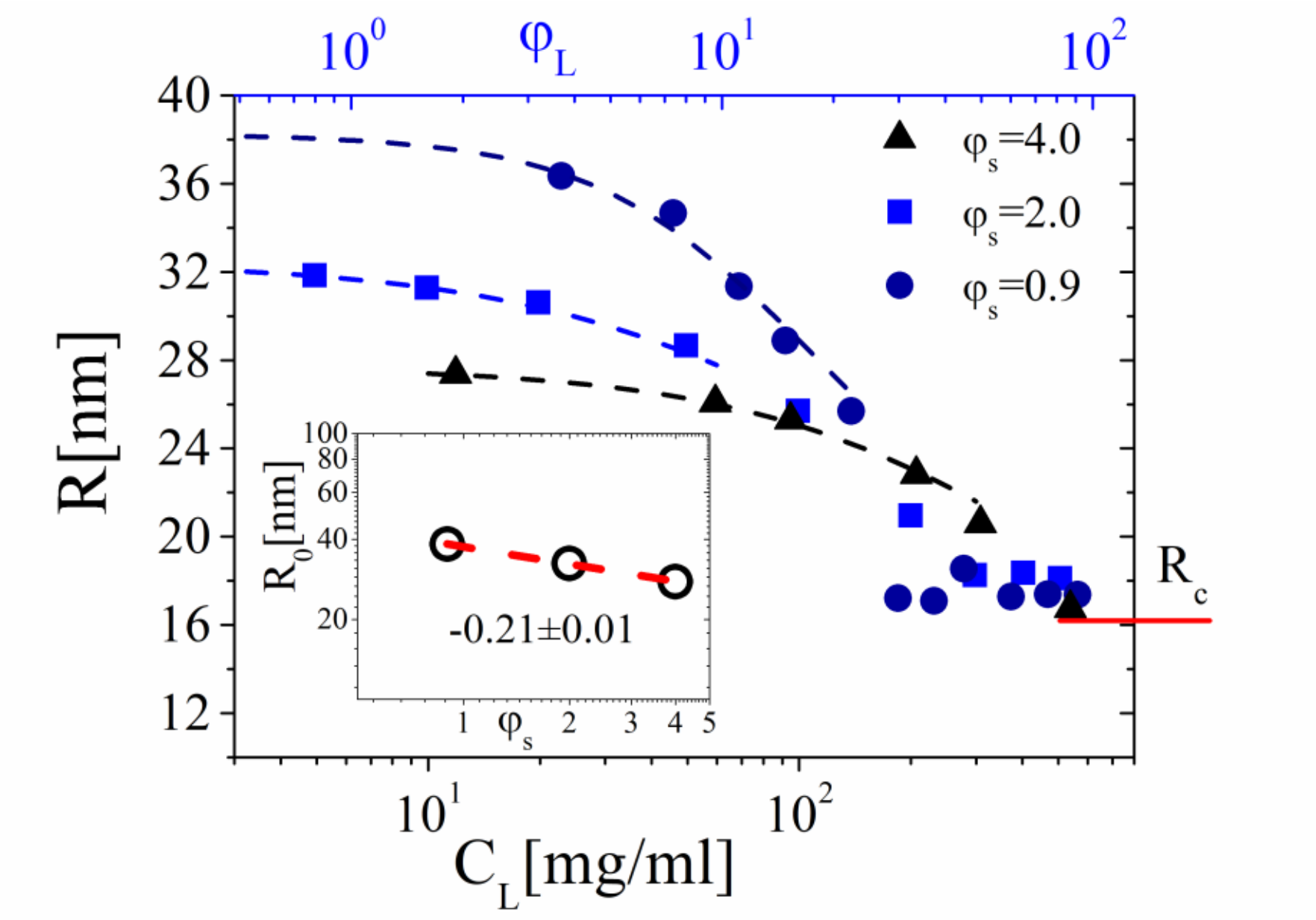

Figure 6. Radius of stars calculated from Equation (1) according to the mapping technique described in Figure 5. The inset shows the unperturbed star radius $R_0$ extracted by fitting the data (dashed line in the main panel) with Equation (5).

Three important characteristics are observed for $R(C_L)$: i) $R_0$ decreases for increasing values of $\varphi_s$; this is expected since every star polymer is expected to undergo osmotic deswelling, even in the absence of free chains, due to the presence of neighboring stars.[53,59] ii) As $\varphi_s$ increases, the shrinkage of the stars caused by the L1000 matrix is reduced, and the onset of the reduction of the star size seems to occur at higher L1000 concentrations, which points to the role of star polymer crowding on the effectiveness of free chain osmotic forcing. More precisely, by increasing $\varphi_s$, the number of chains per star decreases for a set $C_L$ value, hence more chains per unit volume (a higher $C_L$) are needed to induce star collapse: star de-swelling becomes increasingly dominated by star–star repulsions rather than star–chain repulsions. iii) At high $C_L$, $R(C_L)$ appears to converge to the same value at all $\varphi_s$ investigated, very close to $R_c = 16.3$ nm obtained from Equation (2), confirming that the proposed mapping is quantitatively correct and corroborates the hypothesis of the eventual complete collapse of the stars. The emerging scenario of terminal polymer dynamics driven by the configurational space available for the linear chains thus seems to be a good first approximation.

Alternately, it is possible to use mean-field arguments to calculate $R(C_L)$ for low linear chain concentrations. By expanding the osmotic pressure in powers of $C_L$ up to the second order, $\Pi(C_L) = AC_L+BC_L^2$, the size of the stars can be extrapolated in the limit of $C_L = 0$. For an isolated star, the force balance equation obtained by minimizing the single star free energy[28,58] reads as

$$K_0 R + K_1 R^2 \Pi(C_L) - \frac{K_2}{R^4} = 0 \tag{4}$$

where the constants $K_0$, $K_1$ and $K_2$ are functions of the star polymer characteristics (functionality, arm molecular weight, Kuhn monomer size) and temperature.[28] The three terms appearing in Equation (4) are the entropic spring-like force exerted by the star, the osmotic force exerted by the linear polymer matrix, and the excluded volume force that precludes the complete collapse of the star (R = 0), respectively. Knowing that $R_0=R(C_L=0) = (K_2/K_0)^{1/5}$, the above equation can be rearranged to obtain the following approximate expression for the star polymer radius R,

$$R(C_L) \approx \left( \frac{1}{\frac{K_0}{K_1}\Pi(C_L) + \frac{1}{R_0{}^6}} \right)^{1/6} = \left( \frac{1}{(\alpha_1 C_L + \alpha_2 C_L{}^2) + \frac{1}{R_0{}^6}} \right)^{1/6} \tag{5}$$

where $\alpha_1$ and $\alpha_2$ and $R_0$ are constants to be determined by fitting the experimental data.

The best fits obtained using Equation (5) are depicted in Figure 6. In addition, the variation in $R_0$ with $\varphi_s$ is shown in the inset of Figure 6 for three effective volume fractions. The latter plot confirms that the star shrinkage due to other stars is more severe than predicted for linear chain solutions,[58] namely $R(C_L) \sim C_L^{-1/8}$.

This can be attributed to the fact that high functionality stars are more efficient osmotic compressors than linear chains because their osmotic pressure increases with functionality *f*.[60] At the same time, a high-functionality star has a larger elastic modulus compared to a linear chain of equal molecular weight.[15] The net result of star-linear mixing is then a subtle balance between the colloid compressibility and its ability to exchange momentum with the environment, the latter giving rise to the measured osmotic pressure of the suspension. It should be noted that a more pronounced dependence of the osmotic shrinkage of the stars on concentration is predicted by the Daoud-Cotton model for high functionality stars[40] in a good solvent ($R \sim \varphi_s^{-3/4}$) within a concentration range close to the star overlap $C_s^*$ ($C_s^* < C_s < f^{2/5} C_s^*$). This regime has not been observed previously, as most experiments probing directly the star size or diffusion coefficients related to the fast cooperative diffusion, slow self-diffusion, and intermediate structural mode, have been performed with low-functionality stars,[58,61,62] for which this regime can be hardly detected. At any rate, measurements with the S362 stars over a wide concentration range would be needed to validate such a scenario in the present case. It is noteworthy that the value of $R_0$ obtained for $\varphi_s = 0.9$ agrees well with the hydrodynamic radius obtained for single stars in the dilute regime. This provides further support for the validity of the proposed mapping approach, based on the hypothesis that the chain dynamics are controlled by the linear–linear entanglements within the investigated concentration range. It should be also pointed out that around the overlap concentration, star polymers are expected to exhibit crystalline ordering, as a consequence of the increased osmotic pressure arising from the linear chains, due to the inherent non-uniform segment density distribution of the stars.[63,64] Crystalline ordering was not observed in the current system, and a detailed investigation of this phenomenon goes beyond the scope of this work.

The $C_L$ dependence of the plateau modulus $G_p$ of the mixtures is depicted in Figure 7. $G_p$ was consistently selected to match the value of G′ at 100 rad/s. Pure linear polymer solutions exhibit the expected power law dependence[53,65,66] $Gp \sim C_L^{2.3\pm0.1}$ in a good solvent. Coherently with the slowing-down of its dynamics ($\tau_d$), the mixture exhibits remarkable stiffening as the star polymer content is increased at low $C_L$ values, corroborating the scenario of a smooth crossover from bulk to confined dynamics for the linear chains.

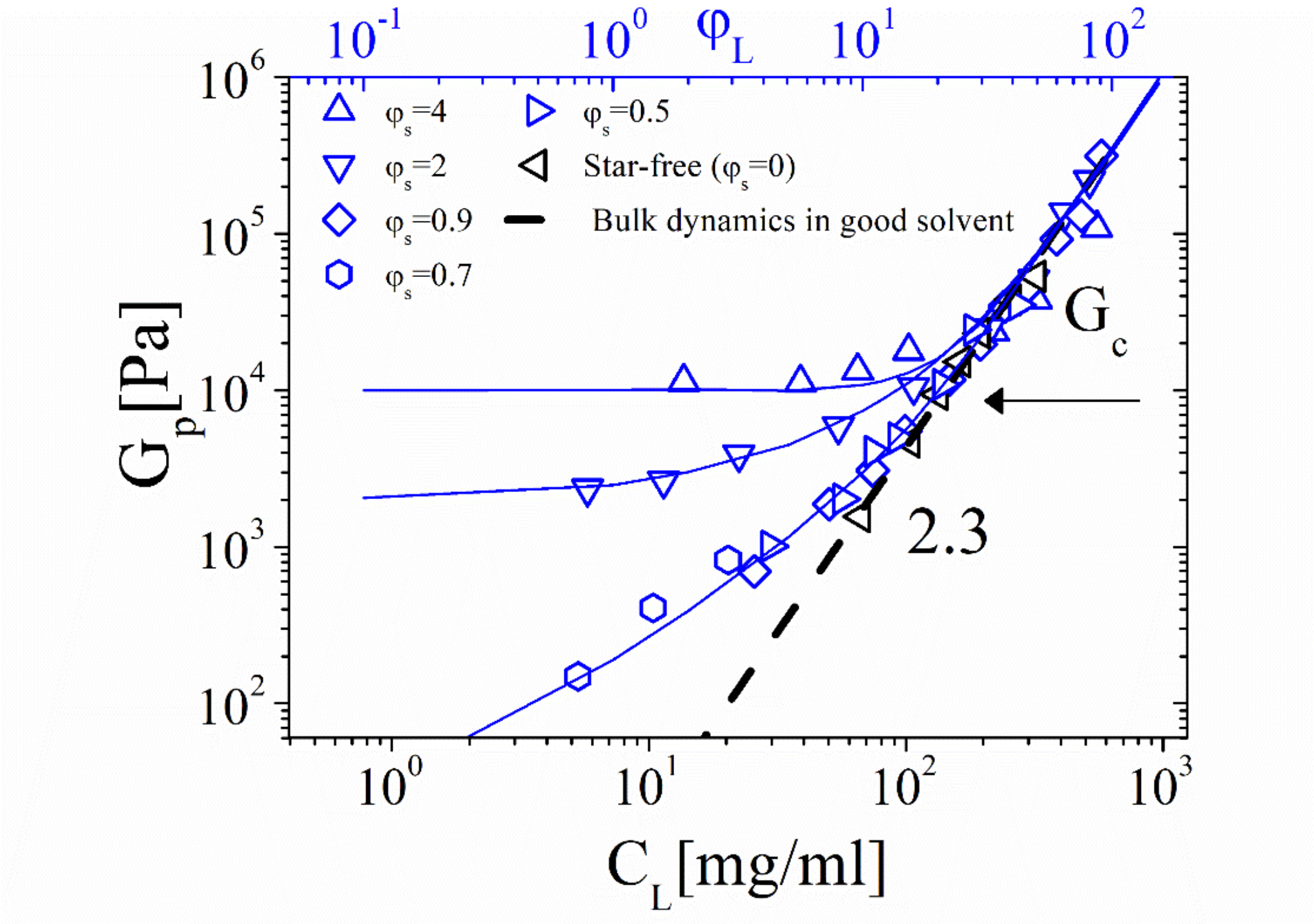

Figure 7. Plateau modulus $G_p$ for star–linear polymer mixtures as function of L1000 concentration $C_L$ at different S362 effective volume fractions $\varphi_s$. The black dashed line corresponds to the power law scaling $Gp \sim \varphi_L^{2.3}$ predicted for semidilute linear polymer solutions. The horizontal arrow points to the critical modulus $G_c$ calculated from Equation (11). Blue solid lines are drawn to guide the eye. The top x-axis shows the effective volume fraction of the linear chains.

Given the phenomenologically common behavior observed for $\tau_d$ and $G_p$, the characteristic linear polymer concentration below which $\tau_d$ and $G_p$ deviate from the scaling characterizing pure (entangled) linear polymers will now be examined. A simple geometric argument is proposed hereafter, based upon the competition between the mesh size of the linear polymer network and the confinement length dictated by the mean distance between stars, to estimate a critical plateau modulus $G_c$ below which a non-negligible influence of star polymers on the modulus of the mixtures is expected (and hence, deviation from linear polymer scaling). First, fluctuation-dissipation scaling can be considered as a correct estimation of the plateau modulus for an entangled linear polymer solution,[6]

$$G_p \approx \frac{k_B T}{\xi(\varphi_L)^3} \qquad (6)$$

where $\xi(\varphi_L)$ is the actual mesh size of the entangled network, which is a decreasing function of the linear polymer effective volume fraction $\varphi_L$.[6,53] The linear polymer high- and low-concentration regimes, illustrated in Figure 7, are defined by the conditions

$$\xi(\varphi_L) << d(\varphi_s, \varphi_L) \qquad \text{High } \varphi_L \qquad (7)$$

$$\xi(\varphi_L) >> d(\varphi_s, \varphi_L) \qquad \text{Low } \varphi_L \qquad (8)$$

where $d(\varphi_s,\varphi_L)$ is the average distance between the ideal outer surfaces of star polymers, i.e., the surface of spheres that cannot be penetrated by linear chains diffusing in the mixture.

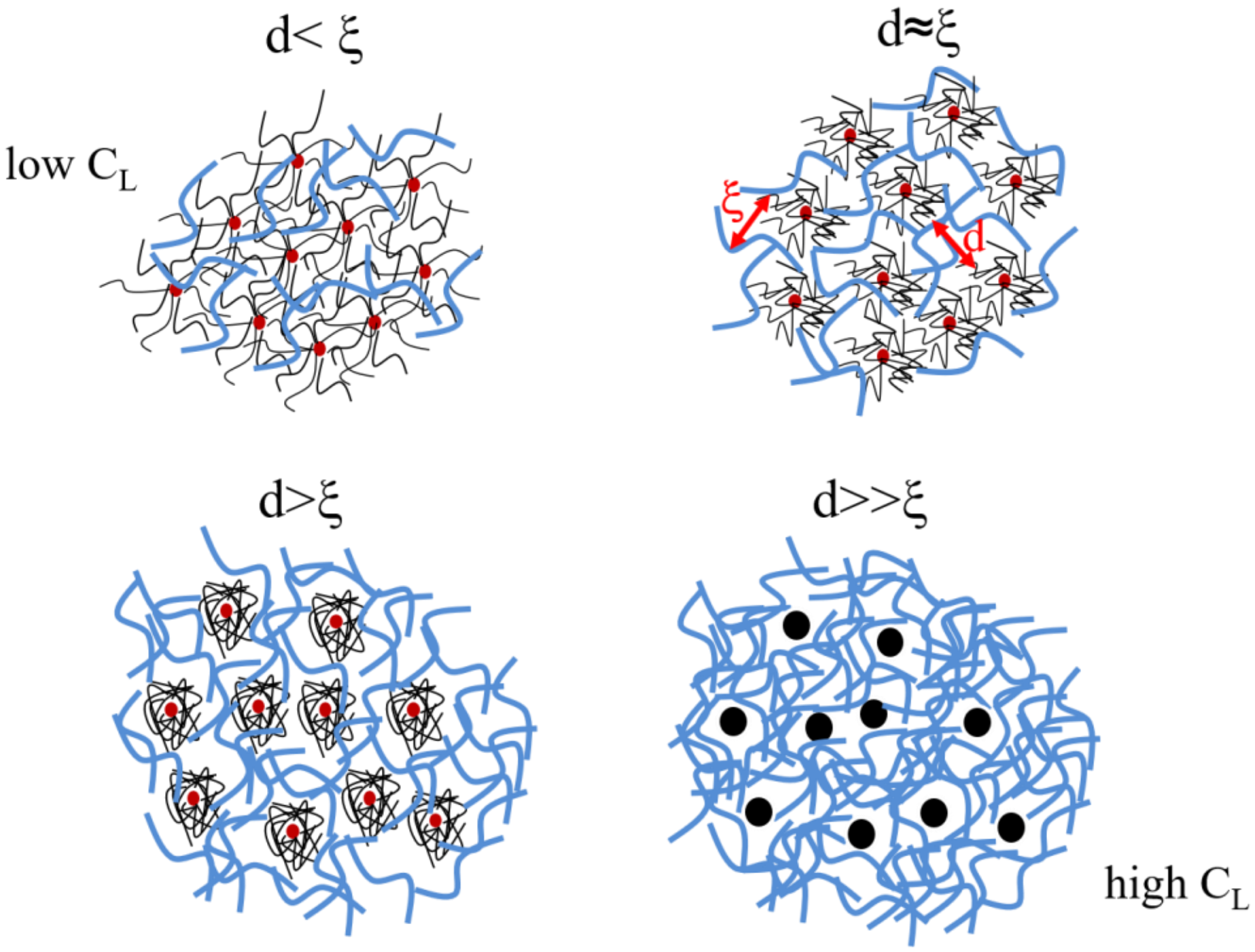

Figure 8. Schematic representation of star–linear polymer mixtures at different linear polymer concentrations $C_L$. Four regimes are shown: i) confined chains (low $C_L$, ξ>d), ii) dynamic

crossover d≈ξ, iii) highly entangled regime (d>ξ), and iv) concentrated regime (d>>ξ) where star polymers are fully shrunk.

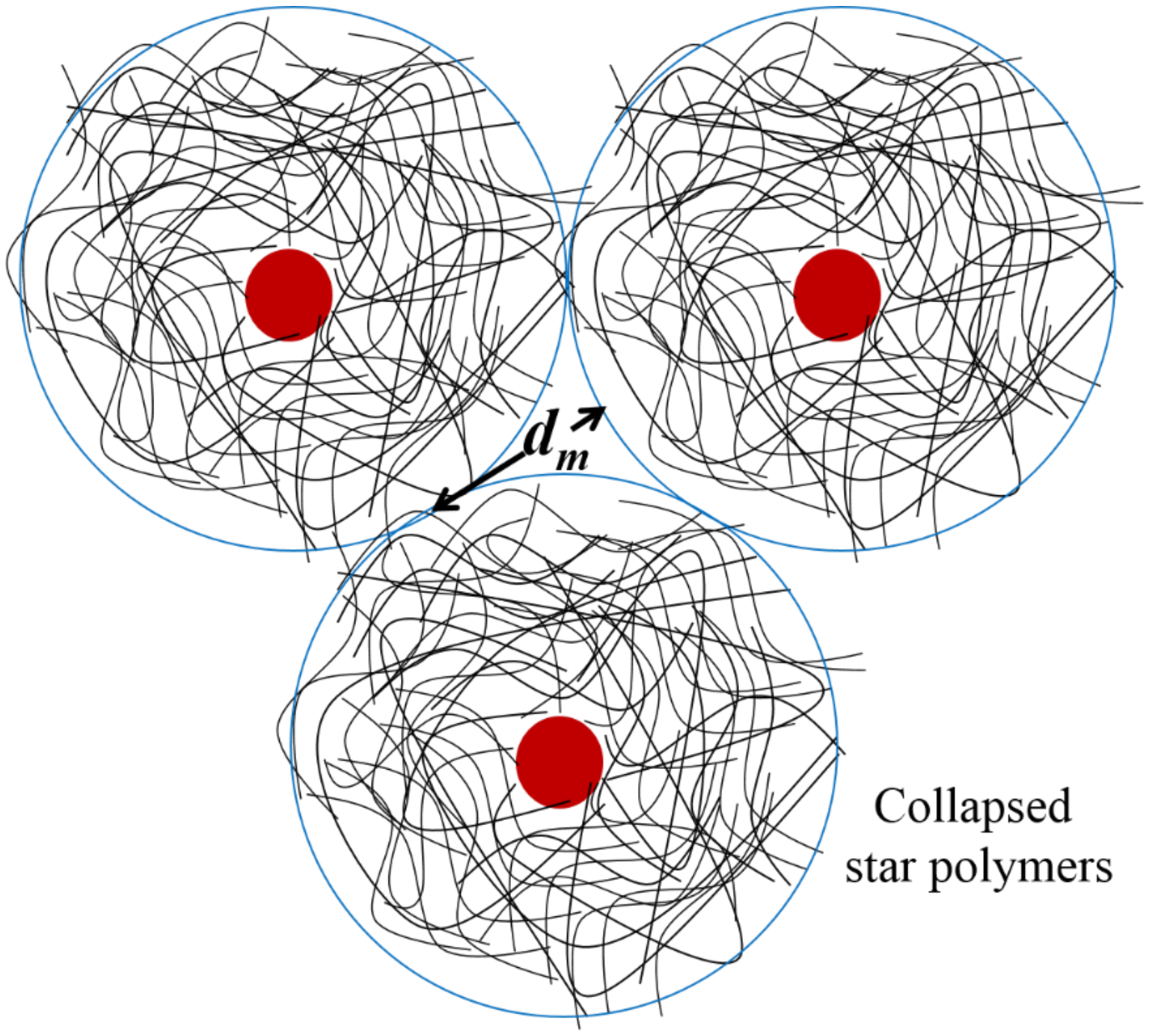

Figure 9. Schematic representation of the minimum confinement produced by three neighboring stars. The parameter $d_m$ is the minimum confinement length affecting the L1000 dynamics when d≤ξ and dictates the crossover between fully bulk and confined dynamics (d ≈ ξ in Figure 8).

As explained above, when the stars reach their collapsed state and expel all the solvent from their interior, it is expected that an effective linear polymer concentration can be found equal to the nominal concentration excluding the stars (Equation 3). This is the reason why, above a certain concentration of linear chains, the convergence of the viscoelastic response ($\tau_d$, $G_p$) of the semidilute polymer matrices of the mixtures towards the behavior of pure linear polymer solutions is observed. At very high $C_L$ and within the investigated range of star concentrations, the mesh size of the linear polymer solution $\xi$ becomes much smaller than the average distance between the outer surfaces of the collapsed stars, i.e. the confinement length *d*. If the system is diluted, by decreasing the amount of linear chains while maintaining a constant mass fraction of stars, a point is reached where the osmotic pressure exerted by the semidilute linear chain solution no longer suffices to compensate for the interaction free energy term (excluded volume) for the stars, which causes their swelling (and wetting) by the molecular solvent. The latter pays a much lower entropic cost (as compared to the linear chains) to penetrate the stars. At that point, the linear chains have access to a volume that is smaller than the nominal volume ($V_{solvent}+V_{chains}$), together with the solvent that wets the inside of the stars. Hence, a deviation is expected from the dynamics (and the scaling) of pure linear chains in good solvent. However, in addition to star collapse upon increasing linear chain concentration, a further necessary condition for the convergence of the dynamics of linear polymers in the mixtures to their classical bulk solution scaling, is that the minimum confinement length in the mixtures is larger than the mesh size $\xi$ of the bulk semidilute linear polymer matrix (Figure 4). Indeed, it is possible to have mixtures at very high star concentrations where the confinement length imposed by the collapsed stars stays always lower than the bulk mesh size of the linear polymer matrix. In such case a complete crossover of the dynamics will not occur.

So, we can write a further condition characterizing the confined-to-bulk dynamics crossover corresponds to $\xi = d_m$,, where $d_m$ is the smallest confinement length in the system and is given by the minimum distance between the surfaces of three close-packed spheres (collapsed stars) in Figure 9 and reads

$$d_m = R_c(\sqrt{3}-1) \qquad (9)$$

where $R_c$, the radius of the collapsed stars (Equation 2), can also be expressed as the radius of a sphere containing $fN_a$ close-packed monomers, i.e.

$$R_c = \frac{1}{2} b (f N_a)^{1/3} \quad (10)$$

By imposing the condition $\xi = d_m$ to fluctuation-dissipation scaling (Equation 6), the following expression is obtained for the critical plateau modulus $G_c$ of the chains marking the crossover from bulk to confined dynamics:

$$G_c = \frac{4 k_B T}{b^3 f N_a (3\sqrt{3} - 5)} \quad (11)$$

Using the known values $T$ = 293 K, $f$ = 362, b = 0.4 nm and $N_a$ = 442, $G_c$ = 8050 Pa is obtained.

It is important to point out that the S362 effective volume fraction $\varphi_s$ does not play any role in Equation (11). This is a limitation of this approach, which restricts the validity of such result to low $\varphi_s$, where both the modulus and the relaxation time are in the low $\varphi_L$ regime (where confined dynamics are observed), and are respectively lower than $G_c$ and its corresponding critical disentanglement time ($\tau_d^c$). In using Equation (11) to obtain the threshold modulus for the dynamics crossover, it is implicitly assumed that complete S362 de-swelling occurs before the mesh size of the L1000 matrix becomes smaller than the confinement length. This condition is fulfilled only at low S362 mass fractions. For this reason, in the limit of very high $\varphi_s$ it is not expected that Equation (11) is valid, since the influence of osmotic de-swelling of the stars upon increasing the L1000 concentration becomes marginal: osmotic de-swelling no longer determines the location of the crossover, the latter occurring presumably at much higher concentrations $\varphi_L$ (and moduli), as suggested by the radii R shown in Figure 6. This high-confinement regime, where it is speculated that non-monotonic behavior of $G_p$ and $\tau_d$ would be possible due to star de-swelling, will be the subject of future investigations. Therefore, it can be stated that Equation

(11) holds true for all mixtures for which confinement never gives rise to $G_p$>$G_c$ in the low $\varphi_L$ regime, and corresponds to a threshold star concentration above which star de-swelling no longer drives the dynamics (it mainly affects the bulk-to-confined dynamics transition). Given the above discussion, Equation (11) is valid for suspensions at $\varphi_s$ = 0.9 and $\varphi_s$ = 2.0, while the mixtures at $\varphi_s$ = 4.0 barely violate the restriction limiting its validity, as indeed in that case $G_p(\varphi_L \rightarrow 0) > G_c$.

As shown in Figure 7, the value of the critical modulus obtained using Equation (11) is in fairly good agreement with the crossover observed experimentally, i.e. where the moduli noticeably deviate from the experimental scaling curve for pure linear polymer solutions. Once again, this corroborates the scenario where linear chain dynamics smoothly transition from confined to bulk behavior, converging to the classical linear polymer scaling as $C_L$ progressively increases, and the stars shrink. Equation (11) also allows the estimation of a critical $\tau_d^c$ characterizing the confined-to-bulk crossover of the dynamics. It can be computed by extrapolating the critical concentration $C_c(G_c)$ = 120.7 mg/ml ($\varphi_L$ = 19.50), i.e. the concentration of a pure L1000 solution with a modulus $G_p = G_c$. The value $\tau_d^c = KC_c^{\alpha}$ = 2.24 s is obtained, in fairly good agreement with the time range where the L1000 chains experience the bulk-to-confined dynamics crossover in the mixtures at low S362 contents ($\varphi_s$ = 0.9 and $\varphi_s$ = 2.0) (see Figure 4). At the highest S362 concentrations investigated ($\varphi_s$ = 4.0), the crossover of $\tau_d(C_L)$ occurs at longer times $\tau_d(C_L)$>20 s. This is expected, since the relaxation time of the confined chains far below the crossover of the dynamics (low $\varphi_L$) is larger than $\tau_d^c$, and Equation (11) gives formally an incorrect prediction, as already discussed for $G_p$.

## CONCLUSIONS

In this work, soft colloid–polymer mixtures were investigated in the hitherto unexplored limit of equal sizes. In particular, the influence of star polymers (as model soft colloids) on the viscoelasticity of the mixtures in a good solvent background was examined by focusing on the response of the linear polymers. In this situation, it was shown how to decompose unambiguously the linear viscoelastic response into polymeric and colloidal contributions. At constant star mass fraction, the dependence of both the entanglement plateau modulus and the terminal relaxation on the concentration of the linear chains was found to follow two distinct regimes. First, the existence of unprecedented dependence of the relaxation time and plateau modulus at low linear polymer concentrations was observed. It was much weaker than that predicted for pure linear chains and became even weaker when increasing the mass fraction of stars. Second, at high linear polymer concentrations, a transition to a much stronger dependence was observed, with the time and moduli following the scaling of entangled polymer solutions. Using simple scaling arguments, this transition was identified with the maximum star osmotic compression, and at the same time the star size was inferred using linear chain relaxation as a probe for the interstitial volume surrounding the stars. This suggests that these mixtures represent an excellent paradigm for an unprecedented transition from confined to bulk dynamics in mixtures of linear chains and soft colloids. These results with symmetric mixtures complete the emerging picture of the extremely rich and intriguing behavior of soft colloid–linear polymer mixtures, with new pathways to tailor the rheology of soft composites and to design novel materials, while also offering an alternate way to study aspects of the colloidal glass transition. The proposed analysis, while restricted to relatively low star mass fractions, can serve as framework to understand polymer dynamics under soft confinement in a large variety of soft systems, as its validity spans colloidal volume fractions up to (and above) the glassy regime.

Finally, it is speculated that the investigation of mixtures with larger star polymer volume fractions, where star de-swelling may yield non-monotonic behavior of the moduli and relaxation times of the chains, represents a challenge for future research.

## ACKNOWLEDGEMENTS

We thank J. Marakis for assistance in the early stages of this work, and Prof C. N. Likos for critical reading of the manuscript and for enlightening discussions. Partial support has been received by the EU (ETN-COLLDENSE, H2020-MCSA-ITN-2014, Grant No. 642774) and the J. S. Latsis Foundation (Grant No. 0839-2012). M. Gauthier thanks the Natural Sciences and Engineering Research Council of Canada (NSERC) for financial support.